# An All-optical Proteretic Switch Using Semiconductor Ring Lasers

Azeemuddin Syed$^a$, Mohamad Tafazoli$^b$, Nima Davoudzadeh$^b$ and M. R. Sayeh$^b$

$^a$Center for VLSI and Embedded Systems, International Institute of Information Technology, Hyderabad, 500032 India, email:- syed@iiit.ac.in
$^b$Department of Electrical and Computer Engineering, Southern Illinois University, Carbondale, IL, 62901 USA

### ABSTRACT

Proteresis, a complementary effect to well-known hysteresis, is an interesting phenomenon to be explored. As we know, the hysteric device is a bistable switch whose states depend on the system's history whereas the proteretic device states' depend on the future. Though an all-optical implementation of the hysteric device has been demonstrated but its complementary behavior has yet to be contrived. Therefore, this work presents the design and development of an all-optical proteretic system using semiconductor ring lasers. The design was achieved by providing a constraint feedforward path to a hysteric feedback system. Additionally, the role of various parameters was studied in order to achieve proteresis. It is observed that the proteretic effect can be achieved only for specific conditions otherwise the system behaves as a traditional hysteric device. The experimental results match well with the theoretical Simulink results.

## 1. Introduction

Proteresis is a phenomenon that behaves as reversed hysteresis. Unlike hysteresis in which there is a delay between two switching thresholds, in proteresis, there is an advancement between the two thresholds. In Greek, hysteresis means which comes after, while proteresis, means which comes earlier [1][2][3]. A proteretic device is expected to speed up the system without losing the benefit of noise immunity.

Analogous to hysteresis, proteresis also is an organic phenomenon observed in materials such as Co:CoO core-shell nanoclusters [4], $Cr_2O_3$ coated $CrO_2$ particles [5], and barium stannate titanate ceramics [6]. Drugs, for instance, lorazepam, alprazolam, diazepam [7], nicotine [8], torasemide [9], and alprazolam [10] have also shown the similar effect. This behaviour has also been perceived in natural processes namely single hydrologic events in rivers [11], phosphorus discharge during storm events [12] [13], solute discharge relationship in rivers during storms [14]. Much of a muchness category of systems were earlier used in software for scheduling controller tasks purpose [15]. Nevertheless, the modeling and implementation of this effect has yet to be explored besides a transition accelerator circuit presented earlier [16]. Recently this work attracted several groups globally, as a result an implementation of proteretic behaviour has been proposed as an optical proteretic bistable device [17] [18]. Lately, a CMOS implementation of a proteretic bistable device was also presented [19].

All-optical devices on the other hand, is gaining lots of interest due to its speed upsurge when compared to its electronics counterparts. As a result, various groups globally have been proposing all-optical devices in diverse domains viz. all-optical phase and amplitude regenerator [20], all-optical neural network [21], all-optical neurosynaptic networks [22] etc. With the miniaturization of semiconductor ring lasers, recently many groups have been proposing all optical devices using semiconductor ring lasers (SRLs). For example, an all optical ultrafast random number generation at 1 Tb/s was proposed using a SRL [23], optical microring resonator based JK flip-flop [24], and an all-optical delta sigma modulator [25] [26] has been designed in the past.

In this paper, we propose a proteretic system implemented using SRLs which has been achieved by a feed-forward feedbacked system. The system was developed using two hysteric devices and an inverter. A total of five SRLs were wielded, two for hysteric device each and one for the inverter.

The rest of the paper is organized as follows: Section 2 contains the design and operation of the proposed system. Section 3 discusses the effect of various parameters on the operation of proteresis device. Section 4 presents the semiconductor ring laser implementation of proteretic system and Section 5 talks about experimental design and results thereof.

## 2. Design and Operation

Proteresis behavior was achieved using two bistable devices and a feed-forward loop. The block diagram of proteretic device using bistable devices represented by their transfer functions is shown in Fig. 1. It can also be seen that the first stage hysteric device is inverted and input is feed-forwarded to the second stage. Although Fig. 1 is for non-inverting proteretic device, it can be modified to the inverting device. This can be achieved by replacing the second stage Schmitt trigger with that of an inverting one. These two hysteric devices were implemented using Schmitt triggers terming them as ST-1 and ST-2.

The transfer functions of both the hysteric devices are shown in Figure 2. Let us understand the conditions under which the system behaves as proteretic device. The two threshold values of inverting bi-stable device i.e. ST-1 are denoted as *a* and *b*. The amplitude of the output of ST-1 is denoted as *c*. Also, it can be seen that the thresholds of non-inverting bi-stable device, ST-2, were stowed as *a'* and *b'*.

It has been observed that the system behaves as proteresis only if certain following conditions has been achieved:

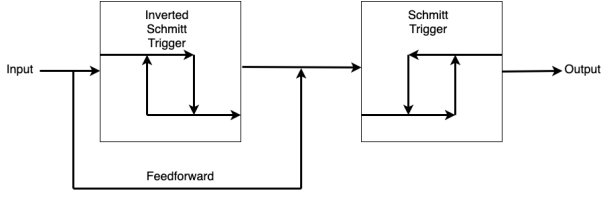

**Figure 1:** A proteretic device using two hysteric bi-stable devices

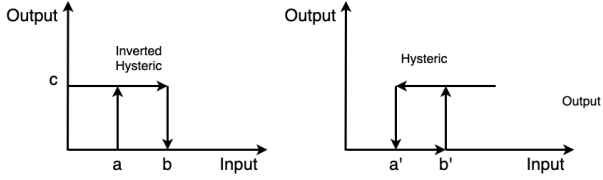

**Figure 2:** Transfer function of inverting and non-inverting Schmitt triggers viz. ST-1 and ST-2

1. The lower threshold value of the second hysteric block ($a'$) should lie between the two threshold values of the first block ($a$ & $b$). Equation 1 defines the condition for lower threshold value of ST-2 in relation with the threshold values.

$$a < a' < b \tag{1}$$

2. The second block's upper threshold value ($b'$) should lie between the sum of output high of the first block ($c$) & it's lower threshold value ($a$) and sum of the output high of first block ($c$) & it's upper threshold value ($b$). Equation 2 defines the condition for upper threshold value of ST-2 in relation with the threshold values and amplitude of ST-1.

$$c + a < b' < c + b \tag{2}$$

3. The hysteric loop width of second block should be less than the output amplitude of the first block ($c$). Equation 3 defines the condition on output strength of ST-1.

$$b' - a' < c \tag{3}$$

The block diagram of proteretic device of Fig. 1 has been implemented in Simulink Matlab which is shown in Fig. 3.

The upper and lower thresholds of ST-1 has been set at $a = 0.2$ and $b = 0.8$ normalized values whereas that of ST-2 were kept at $a' = 0.4$ and $b' = 0.6$. The output value of ST-1 was assigned $c = 0.3$. It can be observed that these values satisfy the conditions mentioned in equations 1-3.

As expected proteresis was achieved with rising edge arriving before the falling edge which can be observed in Figure 4. Results from the Simulink model can be seen in Fig.

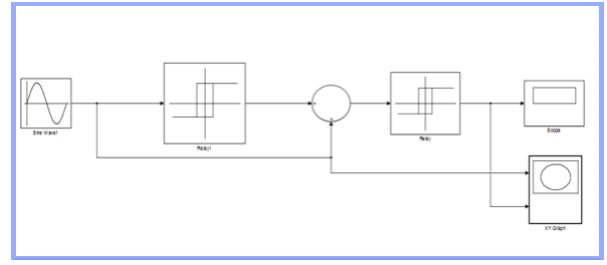

**Figure 3:** A proteretic device using two bi-stable devices

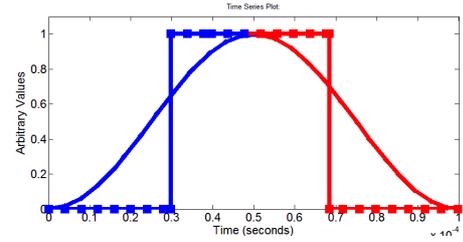

(a) Time domain input and output

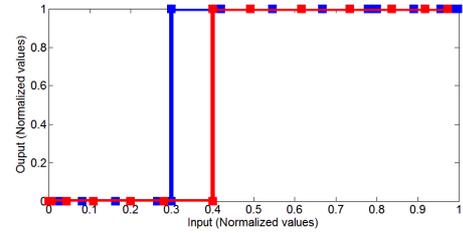

(b) Transfer function

**Figure 4:** Time domain input and output and transfer function of proteretic device simulated on Simulink Matlab

4, shows that red color is rising edge and blue color falling edge.

## 3. Effect of various parameters on operation of proposed system

It would be entrancing to see the effect of each parameter on operation of proposed system. This study would then help the designers to reach optimum values depending on application in which it is used. For example:

1. Effect of varying the amplitude of output of ST-1 i.e. $c$ while keeping the thresholds of ST-1 and ST-2 constant.

2. Effect of varying the threshold values of ST-1 i.e. $a$ and $b$ while keeping the amplitude of its output and ST-2 threshold values constant.

3. Effect of varying the threshold values of ST-2 i.e. $a'$ and $b'$ while keeping the amplitude of ST-1 output and ST-2 threshold values constant.

### 3.1. Effect on proteresis transfer function loop

These effects can be understood by observing the width of loop of the transfer function of the bistable device defined

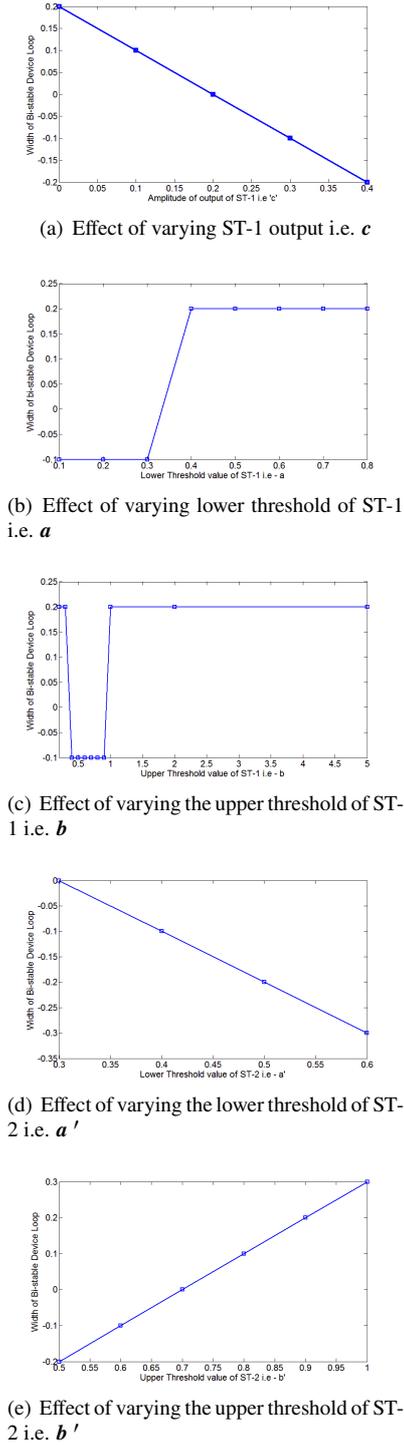

**Figure 5:** Effect of varying various parameters on the width of loop of the bi-stable device

(a) Effect of varying ST-1 output i.e. $c$

(b) Effect of varying lower threshold of ST-1 i.e. $a$

(c) Effect of varying the upper threshold of ST-1 i.e. $b$

(d) Effect of varying the lower threshold of ST-2 i.e. $a'$

(e) Effect of varying the upper threshold of ST-2 i.e. $b'$

as the difference between rising and falling edge threshold values respectively. Therefore, if the width value is positive width, it can be understand that the system is hysteric in nature; and in case it is negative then the loop represents transfer function of a proteretic device. Results of the above study are shown in Figure 5.

From Fig. 5, it can be seen that the bistable loop can be controlled and a larger proteretic loop can be achieved as follows:

1. The value of amplitude of the output of ST-1, $c$, should be kept closer to the value of upper threshold of ST-2, $b'$.

2. The hysteresis loop of the ST-2 should be as minimum (thin) as possible, $a' \approx b'$. This can also be understood from the analytical understanding, the width of proteresis loop $= c - (b' - a')$, hence the maximum width can be achieved if $a' = b'$.

### 3.2. Effect on threshold values

In order to have more control on the system, effect of various parameters on rising and falling edge thresholds of proteretic output was studied. Let $x$ and $y$ be the rising and falling thresholds, respectively. The effect of the five parameters mentioned earlier on $x$ and $y$ are shown in Fig. 6.

Following observations were made:

1. From Fig. 6 (a) and (b), it can be seen that the rising edge threshold of the system decreases linearly with increase in $c$ whereas the falling edge remains constant.

2. From Fig. 6 (c) and (d), the falling edge threshold of bi-stable device switches abruptly with increase in $a$. The rising edge is constant.

3. Both rising and falling edge thresholds change abruptly with increase in $b$ as shown in Fig. 6 (e) and (f).

4. As seen in Fig. 6 (g) and (h), the falling edge threshold of bi-stable system increases linearly with increase in $a'$ and the rising edge is constant.

5. As can be observed from Fig. 6 (i) and (j), the rising edge threshold of bi-stable device increases linearly with increase in $b'$ whereas the falling edge being constant.

## 4. Implementation of proteresis design using semiconductor ring lasers

Semiconductor ring lasers have been used to implement the proposed design of the proteretic device. As seen from the block diagram, it requires two hysteric devices and an inverter as one of the hysteric device is inverted. These bistable hysteric devices can be executed with Schmitt triggers. Two SRLs namely $A$ and $B$ were used to achieve hysteresis effect all-optically, the design of which is shown in Fig. 7. The transfer function can be seen in Fig. 8 where $a$ being the lower threshold and $b$ being the upper threshold.

The output of the Schmitt trigger was then connected to inverter in order to achieve inverted bi-stable device which was named as ST-1. The inverting action was achieved using a single ring laser $E$ as a result, ST-1 was constructed using a total of three SRLs. Its design is shown in Fig. 9; and its transfer function is shown in Fig. 10. It can be seen that the value of output was assigned as $c$ and the threshold values as $a$ and $b$.

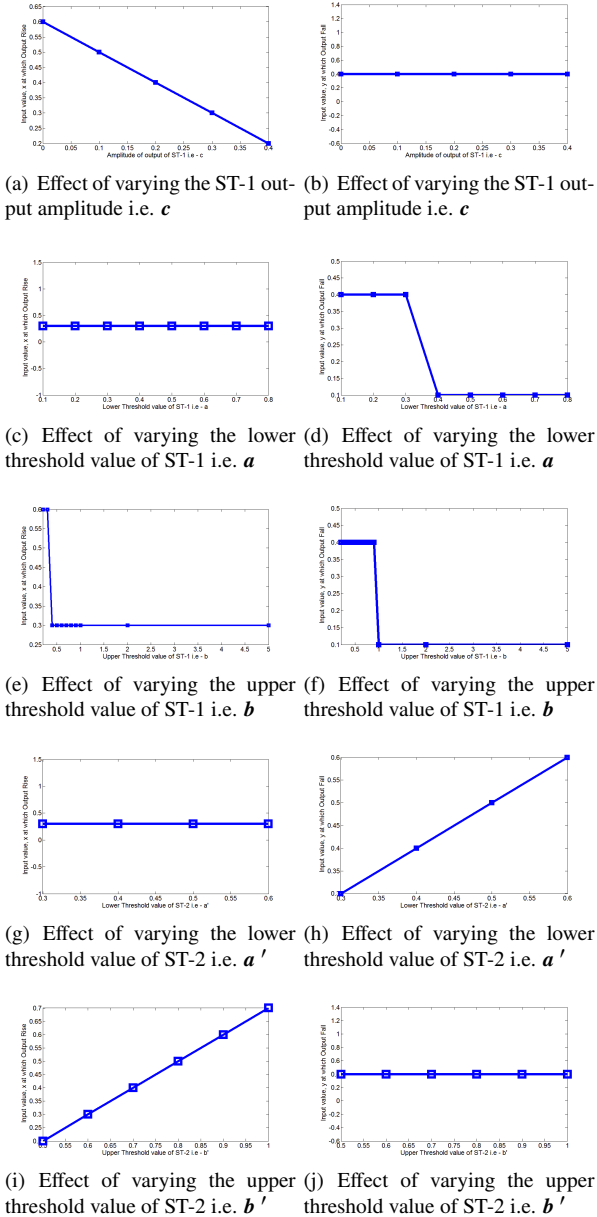

(a) Effect of varying the ST-1 output amplitude i.e. **c**

(b) Effect of varying the ST-1 output amplitude i.e. **c**

(c) Effect of varying the lower threshold value of ST-1 i.e. **a**

(d) Effect of varying the lower threshold value of ST-1 i.e. **a**

(e) Effect of varying the upper threshold value of ST-1 i.e. **b**

(f) Effect of varying the upper threshold value of ST-1 i.e. **b**

(g) Effect of varying the lower threshold value of ST-2 i.e. **a′**

(h) Effect of varying the lower threshold value of ST-2 i.e. **a′**

(i) Effect of varying the upper threshold value of ST-2 i.e. **b′**

(j) Effect of varying the upper threshold value of ST-2 i.e. **b′**

**Figure 6:** Effect of varying various parameters on rising and falling edge threshold values **x** and **y**

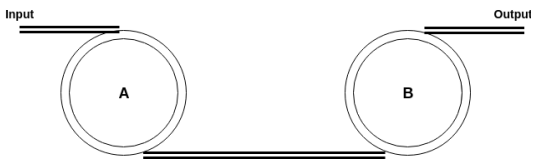

**Figure 7:** Schmitt trigger using SRLs

The second Schmitt trigger (ST-2) with **a′** and **b′** as lower and upper thresholds, respectively was designed similar to that depicted in Figure 7. The SRLs used for this were labeled as **C** and **D**.

Therefore, an inverting bi-stable device (ST-1) has been achieved using three ring lasers; and a non-inverting bi-stable device (ST-2) has been achieved using two ring lasers. The next step is to have control on their threshold values and the amplitude of inverter's output in such a way that the necessary conditions mentioned in equations (1)-(3) are satisfied. The idea is to pre-condition ST-2 with the output of ST-1 in order to achieve proteresis. The control on amplitude and threshold values can be achieved with the help of injection currents of ring lasers [2].

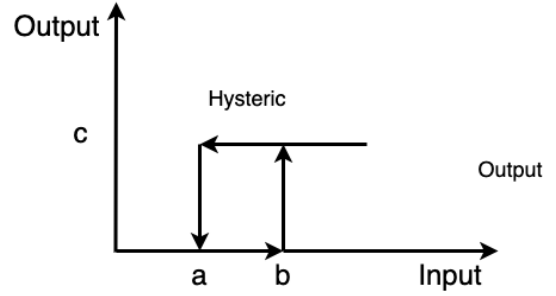

**Figure 8:** Transfer function of Schmitt trigger

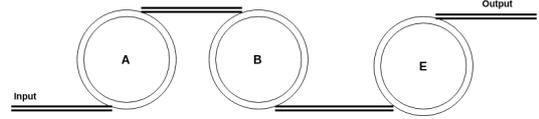

**Figure 9:** Inverted Schmitt trigger

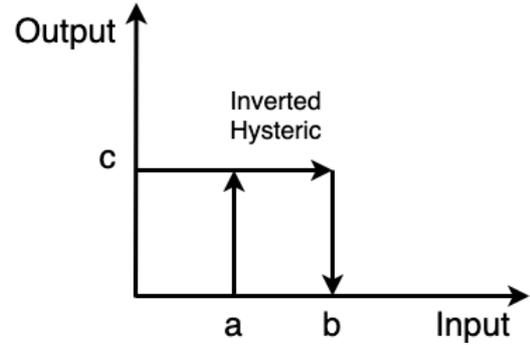

**Figure 10:** Transfer function of inverted Schmitt trigger

As mentioned earlier the width of first hysteric device should be as wide as possible. Hence, in order to increase the width of proteresis loop, following techniques can be applied:

1. The amplitude of the output of ST-1 i.e. **c** should be kept approximately equal to the value of upper threshold of ST-2, **b′**.

2. The hysteresis loop of the ST-2 should be as minimum (thin) as possible, **a′ ≈ b′**.

After analysis of the system, it can be deduced that the width of final proteresis loop is equal to **c** - (**b′** - **a′**), hence the maximum width of the proteresis is equal to the amplitude of output of ST-1 which is achieved at **a′** = **b′**.

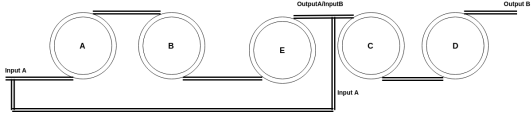

**Figure 11:** Design of proteretic device using five SRLs

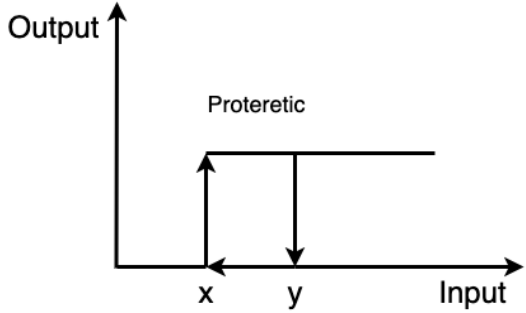

**Figure 12:** Transfer function of proteretic device

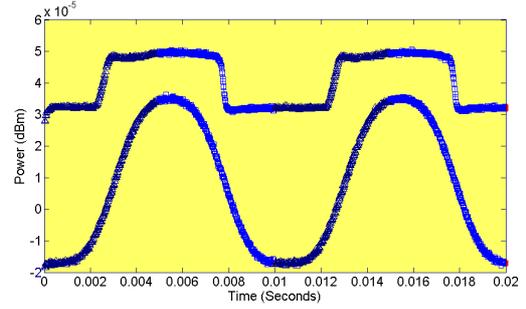

(a) Time domain Signal

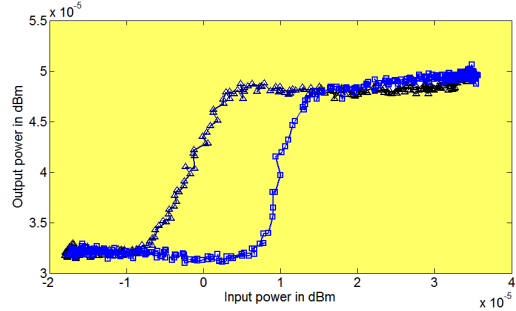

(b) Transfer function

**Figure 13:** Experimental results of proteretic device

After achieving above conditions, the two hysteric devices are connected as shown in the Figure 11. Proteresis has been achieved and; the transfer function of the same can be observed in Figure 12. The new threshold values x and y can now be evaluated as $x = b' - c$ and $y = a'$.

## 5. Experimental Results

The experimental results of the all-optical proteretic device is shown in Fig. 13. The SOA currents of each ring are mentioned below: $I_B = 179$ mA; $I_X = 185.9$ mA; $I_A = 45.7$ mA; $I_C = 271$ mA; $I_Y = 174.5$ mA; and $I_E = 100.2$ mA. A 10 dBm, 1559.91 nm laser is modulated using EOM (electro-optic modulator) with 100 Hz signal.

First Schmitt trigger was implemented employing SRLs B and X as master and slave lasers, respectively; and the second Schmitt trigger was implemented using rings Y and C as master and slave lasers, respectively. SRL A was used as inverter which inverts the output of first Schmitt trigger; and ring laser E was used to control the gain of input given to ST-2.

### 5.1. Estimating range of SOA currents for which the system displays proteresis

In order to estimate the range of permissible SOA currents which effects the threshold values of ST-1, ST-2 and inverter, a study has been done by varying following parameters:

1. Varying the injection current of SOA of SRL A which is similar to changing the amplitude of output of ST-1. It is important to note here that increasing the injection current of SOA in ring laser A decreases the amplitude of ST-1.

2. Varying the injection current of SOA of SRL X which is similar to changing the threshold values of ST-1.

It is important to note here that increasing the injection current of SOA in ring laser X increases the lower threshold value of ST-1.

3. Varying the injection current of SOA of SRL Y which is similar to changing the threshold values of ST-2. It is important to note here that increasing the injection current of SOA in ring laser Y increases the upper threshold value of ST-2.

The ring laser currents are as follows: $I_B = 203$ mA; $I_X = 148.1$ mA; $I_A = 49.6$ mA; $I_C = 321$ mA; $I_Y = 169.2$ mA; and $I_E = 100.2$ mA.

The results obtained thereafter are shown in Fig. 14 and following observations were made:

1. Figure 14 (a) shows the effect of varying SOA current of ring laser A, i.e., output of ST-1 given to ST-2. The current is varied from $I_A = 48.9$ mA to $I_A = 50$ mA. It is observed that proteresis can be achieved in only specific range of currents. It can be seen that proteresis lasts until current value is $I_A = 49.6$ mA; and at $I_A = 50$ mA, the loop changes to hysteresis. It can also be seen that the rise time is not changing much and the falling edge moves from right-side to left-side of the rise time shifting from proteresis to hysteresis. As the current increases from $I_A = 48.9$ mA, the loop gets narrower and finally at $I_A = 50$ mA shifts to hysteresis.

2. Figure 14 (b) shows the effect of varying SOA current of ring laser X, i.e., threshold values of ST-1. The values range from $I_X = 147.9$ to $I_X = 148.4$ mA. Proteresis was achieved at $I_X = 147.9$; and the proteresis

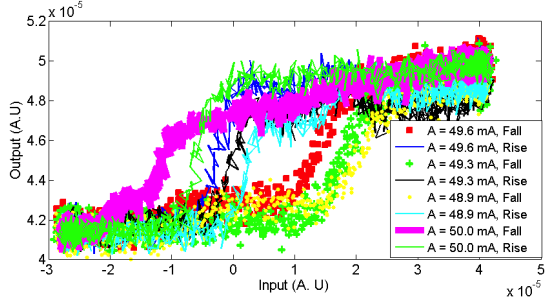

(a) Varying injection current of SRL A, i.e. output of ST-1

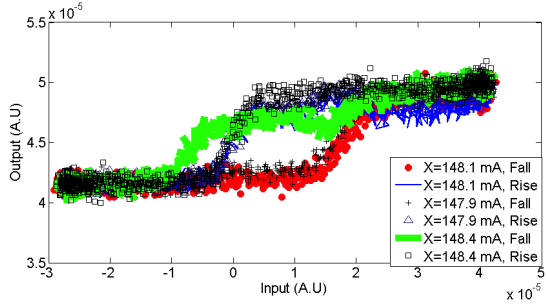

(b) Varying injection current of SRL X, i.e. threshold values of ST-1

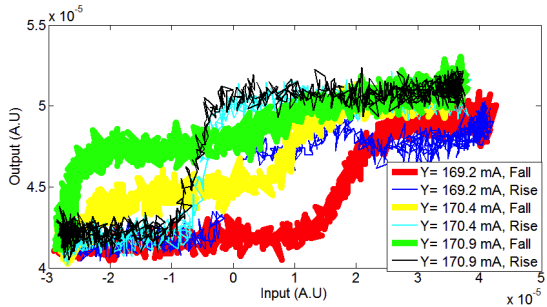

(c) Varying injection current of SRL Y, i.e. threshold values of ST-2

**Figure 14:** Effect of injection current on proteretic loop width

loop changes to hysteresis loop at $I_X = 148.4$ mA. The loop takes multiple stable values if the current is reduced below $I_X = 147.9$ mA.

3. Figure 14 (c) shows the effect of varying SOA current of ring laser Y, i.e., threshold values of ST-2. The values range from $I_Y = 169.2$ mA to $I_Y = 170.9$ mA. Proteresis was achieved at $I_Y = 169.2$; and at $I_Y = 170.4$ mA the proteresis loop changes to hysteresis loop. If the current is increased further to $I_Y = 170.9$ mA, the width of hysteresis loop increases. The loop takes multiple stable values if the current is reduced below $I_Y = 169.2$ mA.

It can be deduced that the proteresis behavior is achieved, where the three conditions mentioned earlier in this report for proteresis are satisfied in a set of very confined values of SOA currents of ring lasers.

The effect of varying injection currents given to SOAs of ring lasers A, X and Y has been studied by plotting the width of bi-stable loop, i.e. the difference between upper and

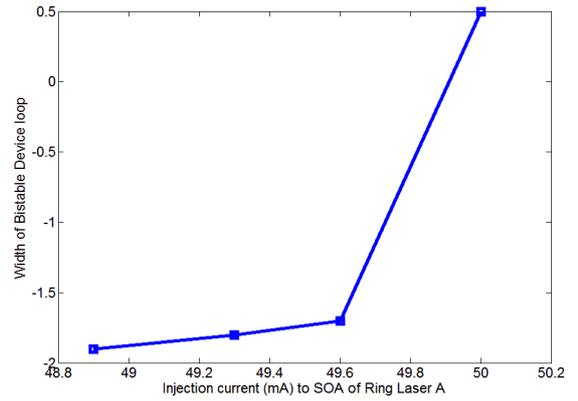

(a) Varying injection current of SRL A, i.e. output of ST-1

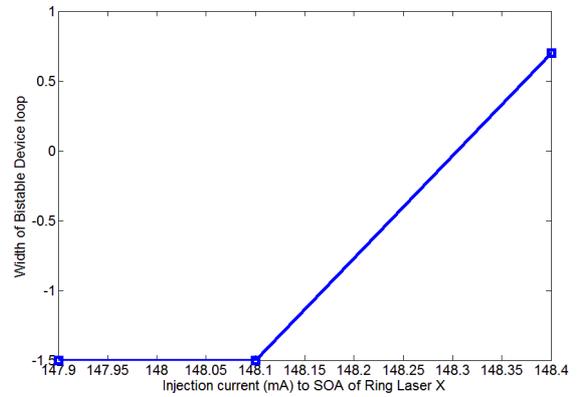

(b) Varying injection current of SRL X, i.e. threshold values of ST-1

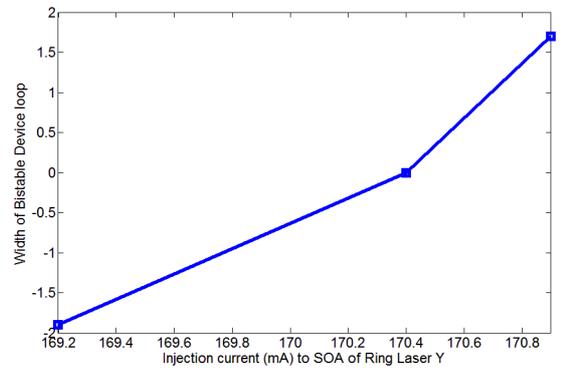

(c) Varying injection current of SRL Y, i.e. threshold values of ST-2

**Figure 15:** Effect of injection current on proteretic loop width

lower thresholds. Basically Fig. 14 is re-plotted showing the width of the loop vs. SOA currents. We hereby refer in Fig. 15, hysteretic loop width as "positive width" and that of proteretic loop as "negative width".

Following observations were made:

1. The experimental result shown in Fig. 15 (a) is in agreement with the Simulink result shown in Fig. 5 (a), i.e. as the amplitude of ST-1 output decreases (by increasing the SOA current of ring laser A), the behavior of bi-stable device changes from proteretic to hysteretic.

2. Similarly result shown in Fig. 15 (b) is in agreement with the Simulink result shown in Fig. 5 (b), i.e. as the lower threshold value of ST-1 increases (by increasing the SOA current of ring laser X), the behavior of bi-stable device changes from proteretic to hysteretic.

3. Result shown in Fig. 15 (c) is in agreement with the Simulink result shown in Fig. 5 (e), i.e. as the upper threshold value of ST-2 increases (by increasing the SOA current of ring laser Y), the behavior of bi-stable device changes from proteretic to hysteretic.

## 6. Conclusion

Semiconductor ring lasers were utilized to design and implement an all-optical proteretic system. A total of five SRLs were required, two for hysteric devices each, and one for the inverter. The effect of various parameters on its operation has been studied and presented. Proteresis is achieved within certain range of threshold conditions of hysteric devices used. This proteretic device can certainly increase the speed of any oscillating or modulating systems.

## References

[1] P. Girard and J. Boissel. Clockwise hysteresis or proteresis. *J. of pharmacokinetics and biopharmaceutics*, 17(3):401–402, 1989.

[2] Younggil Kwon. *Handbook of essential pharmacokinetics, pharmacodynamics and drug metabolism for industrial scientists*. Springer Science & Business Media, 2001.

[3] J Ping Liu, Eric Fullerton, Oliver Gutfleisch, and David J Sellmyer. *Nanoscale magnetic materials and applications*. Springer, 2009.

[4] X-H Wei, R. Skomski, Z-G Sun, and D. J Sellmyer. Proteresis in co: Coo core-shell nanoclusters. *J. of Applied Physics*, 103(7):07D514, 2008.

[5] R. K. Zheng, H. Liu, Y. Wang, and X. X. Zhang. Inverted hysteresis in exchange biased cr 2 o 3 coated cro 2 particles, 2004.

[6] Y. Wei, X.and Feng, L. Hang, S. Xia, Li Jin, and Xi Yao. Abnormal c–v curve and clockwise hysteresis loop in ferroelectric barium stannate titanate ceramics. *Materials Science and Engineering: B*, 120(1-3):64–67, 2005.

[7] Everett H Ellinwood, Douglas G Heatherly, Arlene M Nikaido, Thorir D Bjornsson, and Clinton Kilts. Comparative pharmacokinetics and pharmacodynamics of lorazepam, alprazolam and diazepam. *Psychopharmacology*, 86(4):392–399, 1985.

[8] John W Gorrod and John Wahren. *Nicotine and related alkaloids: absorption, distribution, metabolism and excretion*. Springer Science & Business Media, 2012.

[9] A Paulin, M Schneider, F Dron, and F Woehrlé. A pharmacokinetic/pharmacodynamic model capturing the time course of torasemide-induced diuresis in the dog. *Journal of veterinary pharmacology and therapeutics*, 39(6):547–559, 2016.

[10] MJ Barbanoj, G Urbano, R Antonijoan, MR Ballester, and M Valle. Different acute tolerance development to eeg, psychomotor performance and subjective assessment effects after two intermittent oral doses of alprazolam in healthy volunteers. *Neuropsychobiology*, 55(3-4):203–212, 2007.

[11] M Seeger, M-P Errea, S Beguería, J Arnáez, C Martı, and JM García-Ruiz. Catchment soil moisture and rainfall characteristics as determinant factors for discharge/suspended sediment hysteretic loops in a small headwater catchment in the spanish pyrenees. *Journal of Hydrology*, 288(3-4):299–311, 2004.

[12] Michael J Bowes, William A House, Robin A Hodgkinson, and David V Leach. Phosphorus–discharge hysteresis during storm events along a river catchment: the river swale, uk. *Water research*, 39(5):751–762, 2005.

[13] William A House and Melanie S Warwick. Hysteresis of the solute concentration/discharge relationship in rivers during storms. *Water Research*, 32(8):2279–2290, 1998.

[14] Garnett P Williams. Sediment concentration versus water discharge during single hydrologic events in rivers. *Journal of Hydrology*, 111(1-4):89–106, 1989.

[15] Anton Cervin, Johan Eker, Bo Bernhardsson, and Karl-Erik Årzén. Feedback–feedforward scheduling of control tasks. *Real-Time Systems*, 23(1-2):25–53, 2002.

[16] P. Duchene, M. J. Declercq, and S. M. Kang. Simple cmos transition accelerator circuit. *Electronics letters*, 27(4):300–302, 1991.

[17] N. Davoudzadeh, M. Tafazoli, and M. R. Sayeh. All-optical proteretic (reversed-hysteretic) bi-stable device. *Optics communications*, 331:306–309, 2014.

[18] Mohammad R. Sayeh Mohamad Tafazoli Mehrjerdi and Nima Davoudzadeh Mahboub Sedigh. All-optical proteretic photonic integrated device, 2017. US Patent 9806697.

[19] A. A. K. Reddy, A. Syed, and M. R. Sayeh. A cmos proteretic bistable device. In *2016 IEEE Annual India Conference (INDICON)*, pages 1–4. IEEE, 2016.

[20] J Kakande C. Lundström M. Sjödin P. A. Andrekson R. Weerasuriya S. Sygletos A. D. Ellis L. G. Nielsen D. Jakobsen S. Herstrøm R. Phelan J. O'Gorman A. Bogris D. Syvridis S. Dasgupta P. Petropoulos 'l&' D. J. Richardson R. Slavík, F. Parmigiani. All-optical phase and amplitude regenerator for next-generation telecommunications systems. *Nature Photonics*, 4:690–695, 2010.

[21] Yujun Zhao Yue Jiang You-Chiuan Chen Peng Chen Gyu-Boong Jo Junwei Liu Ying Zuo, Bohan Li and Shengwang Du. All-optical neural network with nonlinear activation functions. *Optica*, 6:1132–1137, 2019.

[22] Youngblood N. Wright C.D. et al. Feldmann, J. All-optical spiking neurosynaptic networks with self-learning capabilities. *Nature*, 569:208–214, 2019.

[23] D. Goulding S. Slepneva B. Kelleher S. P. Hegarty T. Butler, C. Durkan and G. Huyet. Optical ultrafast random number generation at 1tb/s using a turbulent semiconductor ring cavity laser. *Optics Letters*, 41:388–391, 2016.

[24] N. Verma and S. Mandal. Design and performance analysis of optical microring resonator based jk flip-flop. *Optics Letters*, 56, 2017.

[25] M. Tafazoli, N. Davoudzadeh, and M. R. Sayeh. All optical asynchronous binary delta–sigma modulator. *Optics Communications*, 291:228–231, 2013.

[26] A. Syed and M. R. Sayeh. All-optical delta sigma modulator employing semiconductor ring lasers. In *Advanced Photonics Congress*. Optical Society of America, 2012.

Syed Azeemuddin has done his B.E from MJCET, Osmania University in 2003 and MS & PhD from Southern Illinois University Carbondale (SIUC), USA in 2008 and 2008 respectively thereafter joined as faculty member in IIIT-Hyderabad. He has been active in two research areas viz. Radio Frequency Integrated Circuits & Devices and All-Optical Devices using Ring Lasers. He is member of various professional societies - IEEE, OSA, IETE, VLSI Society of India. Along with his teaching and research in these areas he has been active in over-all development of society for eternal success. He is serving community by various methods like Teaching and Conducting workshops on Humanities and Human Values, being reviewer of various conferences and Journals etc. He is recipient of Gold Medal award from MJCET, Osmania University, Masters Fellowship award and Doctoral Fellowship award from SIUC, USA and Visvesvaraya Young Faculty Award from Media Lab Asia, Government of India. Currently he is working on development and fabrication of Integrated Optical Gyroscope, All Optical and RF A/D converters and Modulators, CMOS RF UWB LNA, On-Chip RF Inductors and Bio-Sensors.

M Tafazoli and Nima Davoudzadeh earned his PhD from the Department of Electrical and Computer Engineering, Southern Illinois University Car-bondale.

M R Sayeh is Professor in Department of Electrical and Computer Engineering, Southern Illinois University Carbondale. His main research interests are in the areas of Photonics and Artificial Neural Networks.